\documentclass[useAMS,usenatbib]{mn2e}
\usepackage{aas_macros}
\usepackage{amsmath}
\usepackage{amssymb}
\usepackage{graphicx}
\usepackage[caption=false,lofdepth,lotdepth]{subfig}

\begin{document}

\title[PISNe at reionization]{Pair-Instability Supernovae at the Epoch of Reionization}
\author[T. Pan, D. Kasen, and A. Loeb]{Tony Pan$^1$, Daniel Kasen$^{2,3}$, and Abraham Loeb$^1$\\
$^1$Harvard-Smithsonian Center for Astrophysics, 60 Garden Street, Cambridge, MA 02138, USA\\
$^2$Departments of Physics and Astronomy, UC Berkeley, 366 LeConte Hall, Berkeley, CA \\
$^3$Nuclear Science Division, Lawrence Berkeley National Laboratory}

\pagerange{\pageref{firstpage}--\pageref{lastpage}} \pubyear{2012}

\maketitle

\label{firstpage}
\begin{abstract}
Pristine stars with masses between $\sim$140 and 260 $M_{\odot}$ are theoretically predicted to die as pair-instability supernovae.  These very massive progenitors could come from Pop III stars in the early universe.  We model the light curves and spectra of pair-instability supernovae over a range of masses and envelope structures.  At redshifts of reionization $z\geq 6$, we calculate the rates and detectability of pair-instability and core collapse supernovae, and show that with the \emph{James Webb Space Telescope}, it is possible to determine the contribution of Pop III and Pop II stars toward reionization by constraining the stellar initial mass function at that epoch using these supernovae.  We also find the rates of Type Ia supernovae, and show that they are not rare during reionization, and can be used to probe the mass function at 4-8 $M_{\odot}$.  If the budget of ionizing photons was dominated by contributions from top-heavy Pop III stars, we predict that the bright end of the galaxy luminosity function will be contaminated by pair-instability supernovae.
\end{abstract}
\label{lastpage}


\section{INTRODUCTION}

The life of a massive star ends in a supernova (SN).  The detection of neutrinos from SN 1987A verified the idea that some SNe are set off by the gravitational collapse of the iron core of their progenitor star \citep{Krauss1987}.  However, theory predicts that very massive stars with helium cores between $\sim$64 and 133 $M_{\odot}$ could find another way to blow up, through the thermonuclear explosion of oxygen via the pair-production instability \citep{Rakavy1967, Barkat1967, Heger2002}.  The production of electron/positron pairs in the core softens the equation of state, leading to collapse and the ignition of explosive oxygen burning.  The subsequent thermonuclear runaway reverses the collapse and ejects the entire star, leaving no remnant behind.  Unlike iron core-collapse supernovae (CCSNe), which involves poorly constrained physical processes such as turbulence, pulsations, perhaps rotation and magnetic fields, the physics involved in pair-instability supernovae (PISNe) is fairly well understood and can be modeled with fewer uncertainties \citep{Langer2009}.

Due to the extremely large stellar mass required, the progenitors of
PISNe are expected to be rare, and may only form under unusual
conditions.  One such condition existed in the early universe, when
metal-free Population III stars were born \citep{Loeb2010}.  In star
formation, it is the accretion process that ultimately sets the final
mass of a star.  From dimensional arguments, the mass growth rate is
simply given by the Jeans mass $M_{J} \sim c_{s}^3 G^{-\frac{3}{2}}
\rho^{-\frac{1}{2}}$ over the free-fall time $t_{ff} \sim
1/\sqrt{G\rho}$, implying $dM/dt \propto c_{s}^3/G \propto
T^{\frac{3}{2}}$, where the sound speed $c_{s} \sim \sqrt{kT/m_{p}}$.
In present day star-forming regions, heavy elements radiatively cool
the gas to a temperature as low as $T\sim 10$K.  However, in
primordial clouds, the primary coolant at low temperatures is
molecular hydrogen, which can only cool the gas to $T\sim 200-300$K,
implying an accretion rate higher than present day by two orders of
magnitude.  Hence, theoretical studies suggest that the initial mass
function (IMF) of Pop III stars might have been biased toward masses
much higher than today, e.g. several hundred $M_{\odot}$
\citep{Bromm2004}.  The nucleosynthesis imprints of this top heavy IMF
have been seen in globular clusters and damped Lyman alpha systems
\citep{Cooke2011, Puzia2006}.  Moreover, massive stars have strong
winds driven by radiation pressure through spectral lines, with a mass
loss rate scaling with stellar metallicity $\dot{M}\propto Z^{0.5\sim
0.7}$ \citep{Vink2001, Kudritzki2002}.  Most PISNe should therefore be
from Pop III stars, which have weak radiation-driven winds due to their
extremely low metallicities, and retain enough of their initial masses
at the end of their lives to undergo a pair-instability explosion.

Naturally, studies of the rates and detectability of PISNe 
focused on high redshifts before reionization.  \citet{Mackey2003}
found the PISNe rate to be $\sim 50$ deg$^{-2}$ yr$^{-1}$ at $z > 15$,
while \citet{Weinmann2005}, using more conservative assumptions for
the number of PISNe produced per unit Pop III stellar mass formed,
found the PISNe rate to be $\sim 4$ deg$^{-2}$ yr$^{-1}$ at similar
redshifts.  Assuming that only one supermassive Pop III star forms in
unenriched minihalos, and that none form in protogalaxies,
\citet{Wise2005} found the PISNe rate be $\sim 0.34$ deg$^{-2}$
yr$^{-1}$ at $z\sim 20$.  After our paper was submitted,
\citet{Hummel2011} presented a complimentary analysis of the source
density of PISNe from pristine minihalos, and determined the
observability of such events with the \emph{James Webb Space
Telescope} (JWST), finding approximately $\sim 0.4$ PISNe visible per
JWST field of view at any given time.  PISNe after the epoch of
reionization were also considered; \citet{Scannapieco2005a} calculated
a suite of PISNe model light curves with blackbody spectra, and
analyzed the detectability and rates of PISNe from Pop III stars
formed from leftover pristine gas at $z \lesssim 6$.  As for CCSNe
during reionization, \citet{Mesinger2006} presented detailed
predictions for the number of core collapse SNe that JWST could
observe as a function of different survey parameters.

In this paper, we present light curves and spectral time series for
PISNe from our multi-wavelength radiation-hydrodynamics simulations.
As the stellar population responsible for reionization is currently
unknown, instead of predicting a fixed SNe rate, we normalize the star
formation rate by requiring that enough ionizing photons must be
produced by either Pop III or Pop II stars in protogalaxies to
complete reionization by $z \sim 6$, and calculate the rates of
pair-instability, core-collapse, and Type Ia SNe and their
detectability with JWST for these two scenarios; the actual SNe rates
will be in between these limiting cases.  We show that using the
observed rates of these SNe, it is possible to distinguish the
contribution of Pop III and Pop II stars toward reionization by
characterizing the IMF at that time.

\section{LIGHT CURVES AND SPECTRA}
The stellar evolution and explosion of PISN models, and the resulting broadband light curves and spectral time-series are described in detail in \citet{Kasen2011}; here we summarize the results.  Models R150, R175, R200, R225, R250 represent explosions of 150-250  $M_{\odot}$ red supergiant stars, respectively, each with their hydrogen envelope intact.  In principle, blue supergiants  are also possible progenitors of PISNe, but convective mixing of metals into the hydrogen envelope makes it more likely that the progenitor dies as a red supergiant.  Models He80, He100, He130 were explosions of 80, 100, 130 $M_{\odot}$ bare helium cores.  Such models may represent stars that lost they hydrogen envelope  due to a prior pulsational phase or through binary interactions.  Here we use an approximate empirical relation between the helium core mass and the progenitor main-sequence mass \citep{Heger2002}:
\begin{equation}
M_{He} \approx \frac{13}{24}(M_{ZAMS}-20M_{\odot}).
\end{equation}
Properties of all presupernova stars and their explosions are given in Table \ref{ProgenitorModelTable}.

\begin{table}
 	 \caption{Parameters of supernova explosion models.  The R-prefix models refer to red supergiant progenitor PISNe, and the He-prefix models refer to the exposed helium core PISNe.  The proxy core-collapse SN model (CC), a model Type~IIP supernova, is shown for comparison.  $R_0$ is the presupernova radius.  $M_i$, $M_f$ are the initial and final masses of the progenitor, respectively, while $M_{He}$ is the helium core mass and $M_{Ni}$ is the amount of $^{56}$Ni synthesized in the explosion.  All mass units are in $M_{\odot}$.}
	 \begin{tabular}{ | l | l | l | l | l | l |}
    \hline
    Name		& $M_i$ 	& $M_f$ 	& $M_{He}$ 	& $M_{Ni}$	& $R_0$ ($10^{12}$cm)\\ \hline
    R150 	& 150		& 142.9 	& 72.0 		& 0.07		& 162						\\ 
    R175 	& 175		& 163.8 	& 84.4 		& 0.70		& 174						\\ 
    R200 	& 200		& 181.1 	& 96.7 		& 5.09		& 184						\\ 
    R225 	& 225		& 200.3 	& 103.5 		& 16.5		& 333						\\ 
    R250 	& 250		& 236.3 	& 124.0 		& 37.86		& 225						\\ \hline
    He80 	& 80		& 80	 	& 80 			& 0.19		& -						\\ 
    He100 	& 100		& 100 	& 100 		& 5.00		& -						\\ 
    He130 	& 130		& 130 	& 130 		& 40.32		& -						\\ \hline
    CC 		& 15		& 13.3 	& - 			& 0.28		& 44						\\ \hline
    \end{tabular}
\label{ProgenitorModelTable}
\end{table}

A few days after the explosion, hydrodynamical processes subside and the ejected material reaches a phase of nearly free expansion.  The energy powering the subsequent light curve may derive from three possible sources: (\emph{i}) Lingering thermal energy from the explosion itself; (\emph{ii}) The radioactive decay of synthesized $^{56}$Ni; (\emph{iii}) The interaction of the ejecta with a dense circumstellar medium.  Thermal energy suffers adiabatic losses on the expansion timescale $t_{ex} = R_{0}/v$, and so source (\emph{i}) is only significant for stars with large initial radii $R_{0}$.  Circumstellar interaction has not been included in the models discussed here.

We have computed light curves and spectral time series of the explosion models using the time dependent radiative transfer code SEDONA \citep{Kasen2006}.  All models shown here assume spherical symmetry, and calculations of atomic level populations assume local thermodynamic equilibrium.  Using Monte Carlo methods, we solve the full multi-wavelength radiative transfer problem using realistic opacities as a function of wavelength, composition and temperature, over millions of line transitions.  Unlike previous blackbody models \citep{Wise2005, Scannapieco2005a}, our results allow us to calculate more accurate light curves for any given color bands and to study the time evolution of the supernova colors and spectral features.

\begin{figure}
\centering
\includegraphics[width=1.0\columnwidth]{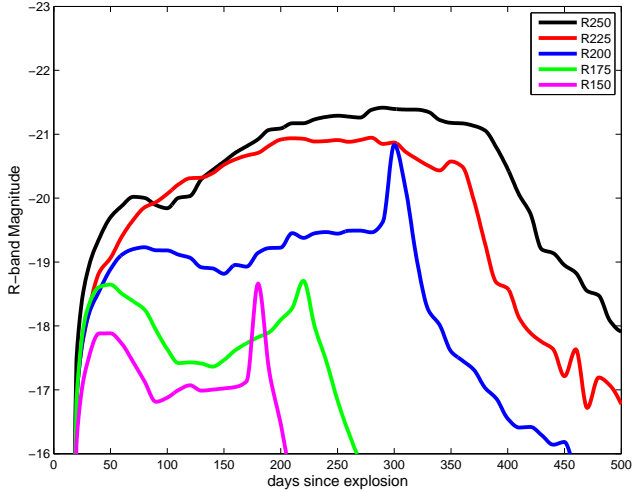}
\caption{Rest frame R-band light curves for the red supergiant progenitor models.  In some models, a brief spike in luminosity occurs at the end plateau when radiation is released by hydrogen recombination.  The sharpness of the spike may be exaggerated by the lack of numerical convergence of the ionization front recession. }
\label{restframeRlightcurve}
\end{figure}

\begin{figure}
\centering
		\includegraphics[width=1.0\columnwidth]{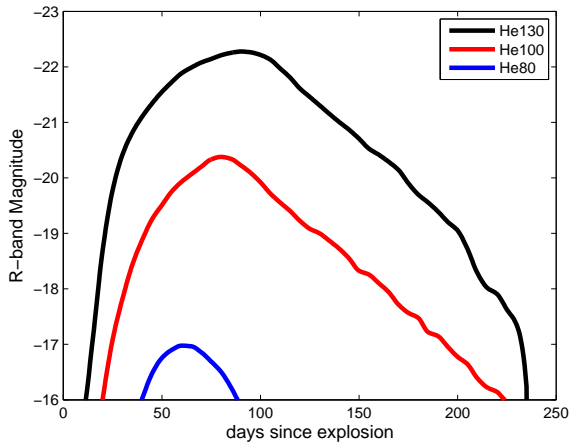}
	\caption{ Rest frame R-band light curves for the helium core progenitor models.}
\label{restframeHelightcurve}
\end{figure}

The shape and duration of PISN light curves depend on the mass and radius of their progenitors.  Model R250 shows a weak and then strong peak in its light curve (Figure \ref{restframeRlightcurve}), the initial peak powered by thermal energy and the second by the radioactive decay of $^{56}$Ni.  The heating from radioactive decay delays the inward-propagating recombination wave from ejecta cooling, regulating the electron scattering opacity (and thus the release of thermal energy), and causing the second peak to rise at 200-300 days, which reaches a spectacular brightness of $\sim -21.5$ mag.  However, model R150 produces very little $^{56}$Ni, and therefore lacks a prominent second peak; the light curve is essentially thermally powered and reaches a brightness less than that of a Type Ia SN.  

The helium core models are more compact and hence lack an initial thermal peak (Figure \ref{restframeHelightcurve}).  Model He130 reaches an exceptional peak brightness of $2\times 10^{44}$ ergs s$^{-1}$, whereas Model He80 demonstrates that despite being massive and energetic, not all PISNe are bright.  This steep mass-luminosity relation for PISNe suggests that to increase the sheer number of SNe detected, it is better to conduct a wide rather than deep survey of the sky \citep{Weinmann2005}.

\begin{figure}
\centering
		\includegraphics[width=1.0\columnwidth]{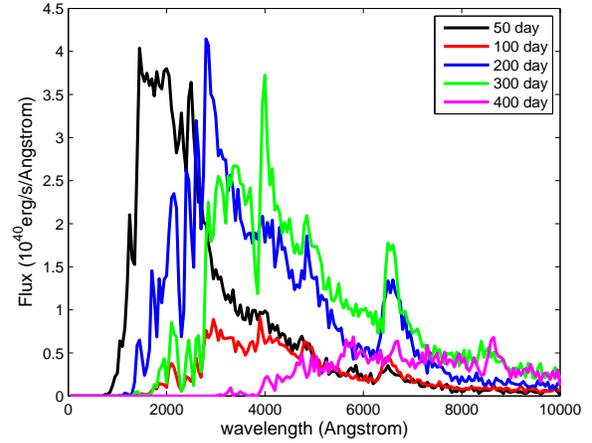}
	\caption{ Time evolution of the rest frame spectra for the R250 red supergiant model.}
\label{restframeR250}
\end{figure}

\begin{figure}
\centering
		\includegraphics[width=1.0\columnwidth]{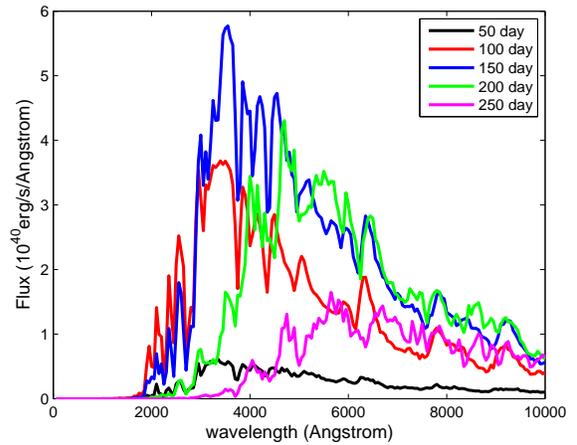}
	\caption{ Time evolution of the rest frame spectra for the He130 helium core model.}
\label{restframeHe130}
\end{figure}

The spectra of a PISN resemble that of average SNe, with P-Cygni line profiles on top of a blackbody, see Figures \ref{restframeR250}, \ref{restframeHe130}.  For the RSG models, at early times, the spectrum is rather featureless with only weak Balmer and calcium lines, reflecting the low abundance of metals in unburned ejecta.  The spectral energy distributions of the models are blue at earlier times ($\la 50$ days) but become redder over time as the expanding ejecta cools.  In addition, line blanketing of the bluer wavelengths becomes more prominent over time, as the photosphere recedes into the deepest layers which are abundant in freshly synthesized iron group elements.  For PISNe at the redshifts of reionization, JWST will mostly be observing in the rest frame UV, so it is important to use more accurate spectral models, rather than the blackbody models of \citet{Scannapieco2005a}.  

Spectroscopic or rest frame UV observations of PISNe may be able to constrain the metallicity of the progenitor star.  However, the hydrogen envelope may be polluted by newly synthesized metals mixed out during the explosion.  \citet{Chen2011} simulated multi-dimensional models of PISNe to predict the degree of mixing.  They found relatively small fluid instabilities generated from burning at the boundaries of the oxygen shell, and concluded that PISNe keep their onion-shell structure in the explosion, until the reverse shock passes which generates Rayleigh-Taylor instabilities.  This is in contrast with CCSNe, in which a shock runs through the inner metal-rich core, inducing the growth of instabilities and mixing.  Also, ordinary Pop II/I CCSNe have non-zero metallicity in their hydrogen envelopes to begin with.  Hence, metal lines in early-time spectroscopy might be able to distinguish PISNe from CCSNe, before the photosphere has receded deep into the ejecta.  With a little mixing, N and possibly some C and O might appear in the early spectra of PISNe, but PISNe will not have any Si, Ni, Fe lines \citep{Joggerst2011}.  This best applies to the red supergiant models, as the Helium core models undergo significant burning and have spectra that show many metal lines at maximum light.

\section{Supernovae During Reionization}

Observations of quasar absorption spectra \citep{Fan2006} indicate
that reionization was completed by $z=6$.  It is believed that most of
the ionizing photons came from stars \citep{Loeb2010, Bouwens2011a}.
Although the very first stars could have ignited as early as $z\sim
30-40$, due to the exponential nature of structure formation, most of
ionizing photons originated from stars born in the later stages of
reionization at $z\sim 10$.  Although it is not known which population
of stars dominated at this epoch, an unusual stellar mass function
10-20 times more efficient than the standard Salpeter IMF in producing
ionizing photons is required at $z\sim 6$ \citep{Cen2010}.  This
favors the existence of a top-heavy Pop III stars at these redshifts,
which may be observable via their extraordinary deaths as PISNe if the
IMF included mostly stars between 140 and 260 $M_{\odot}$.  Moreover,
the observed rates of PISNe, CCSNe, and Type Ia SNe may be used to
infer the IMF responsible for reionization at $z\gtrsim 6$.

\subsection{The Initial Mass Function}

The ionizing photon yield per baryon incorporated into present day stars with a Salpeter IMF is $\overline{\eta_{\gamma}}\sim 4000$. However, if the IMF is dominated by massive metal free stars ($M > 100 M_{\odot}$), then $\overline{\eta}$ can be up to a factor of 20 higher \citep{Bromm2001,Raiter2010}.  The transition from Pop III to Pop II/I star formation is thought to occur at a critical metallicity of $Z_{crit}\sim 5 \times 10^{-4}Z_{\odot}$, above which cooling and fragmentation become efficient, which stops the preferential formation of massive stars \citep{Bromm2003}. 

This transition can be associated with the assembly of atomic H cooling halos with virial temperatures $> 10^4$K \citep{Haiman2009}.  Molecular hydrogen is fragile to photodissociation, and the molecular coolant in halos are likely depleted after a single episode of metal free star formation.  Therefore, molecular hydrogen halos are unlikely to allow continued formation of stars above $Z_{crit}$.  Subsequent star formation only occurs when the deeper gravitational potential wells of atomic H cooling halos are assembled, corresponding to a virial temperature of $T_{vir}\approx 10^4$K and a minimum halo mass of $M_{halo}\approx 10^8M_{\odot}$  The gas in these halos will thus have already gone through a burst of primordial star formation, and contain traces of metals, leading to Pop II star formation.  Most of the photons responsible for reionization will come from the profusion of these Pop II stars in this scenario, although without contribution from Pop III stars, this may require an unrealistic star formation efficiency, see Figure \ref{SFRchart}.

There is another possibility.  Most molecular $H_{2}$ cooling halos may not have formed stars at all, due to global $H_{2}$ photodissociation by an early cosmic background of 11.2-13.6 eV photons (the Lyman-Werner band), to which the universe is otherwise transparent.  In this scenario, the majority of primordial star formation will appear in atomic H cooling halos with $M_{halo}\approx 10^8M_{\odot}$.  During blowouts from repeated SN explosions, these halos allow most of their self-generated metals to be accelerated into the IGM as SN ejecta, but, in contrast to smaller molecular $H_{2}$ cooling halos, these halos hold on to most of their interstellar gas \citep{MacLow1999}, and can have significant Pop III star formation. Coupled with the high ionizing efficiency of massive metal free stars, in this scenario Pop III stars will make a significant contribution to reionization.

Hence, we consider two mutually exclusive scenarios for reionization, where either Pop III or Pop II stars reionize the universe; the actual star formation history of reionization will be in between these limiting cases.  As for the IMF in each scenario, for massive, metal free Pop III stars, we use either a Salpeter IMF slope $dN/d\log M\propto M^{-1.35}$, or a flat IMF slope $dN/d\log M\propto M^0$ hinted by recent simulations \citep{Clark2011, Greif2011}, with $M_{upper} = 500M_{\odot}$ and $M_{lower}=1M_{\odot}$.  Note that the resulting PISN rates are not sensitive to the upper and lower mass bounds of reasonable Pop III IMFs.  As long as $M_{upper} > 260 M_{\odot}$, due to the steepness of $dN/dM$, there are not enough stars at the most massive end to affect the overall normalization.  Moreover, for our SFR model described in the next section, any reasonable $M_{lower}$ ranging from  $0.1$ - $10 M_{\odot}$ makes negligible difference to the PISN rates of Pop III stars.  This is because we normalize star formation by requiring the generation of enough stellar UV photons necessary to reionize the universe, and massive stars $M \gtrsim 10 M_{\odot}$ are drastically more efficient at producing ionizing photons.  In essence, we fix the number of massive stars produced in any IMF, but are free to vary the number of low mass stars, as the latter do not contribute to reionization anyway.  $M_{lower} = 1 M_{\odot}$ was chosen to match the smallest Pop III stars seen in recent simulations by \citet{Clark2011}.

For Population II stars forming with traces of metals, we use a Salpeter IMF with $M_{upper} = 125M_{\odot}$ and $M_{lower} = 0.1M_{\odot}$, where we include a factor 0.7 in the mass integral to account for the reduced number of low mass stars in a realistic IMF \citep{Fukugita1998}, compared to the original Salpeter IMF.  The different IMF models are tabulated in Table \ref{IMFTable}.

\begin{table}
 	 \caption{Model parameters of the different IMFs.  Here $\alpha$ is the slope of the stellar mass function, i.e. $dN/d\log M\propto M^{\alpha}$, and the slope of the Pop II IMF flattens at $M<0.5M_{\odot}$.  All mass units are in $M_{\odot}$.}
	 \begin{tabular}{ | l | l | l | l | l | l |}
    \hline
    IMF model $\phi(M)$	& $M_{lower}$ 	& $M_{upper}$ 		& $\alpha$ 		& $\overline{\eta}_{\gamma}$		\\ \hline
    Pop III Salpeter 	& 1 				& 500 				& -1.35 			& 28683									\\ 
    Pop III Flat 			& 1				& 500 				& 0				& 77087									\\ 
	 Pop II 					& 0.1				& 125					& -1.35			& 5761									\\ \hline
    \end{tabular}
\label{IMFTable}
\end{table}

\subsection{The Star Formation Rate}

We calibrate the SFR by requiring enough UV photons are produced by stars so as to ionize the intergalactic medium (IGM) by the end of reionization.  This requires $C \sim 10$ ionizing photons per baryon in the IGM, accounting for recombinations \citep{Trac2007}.  Using the time-averaged ionizing flux and stellar lifetime for individual stars from \cite{Schaerer2002}, we find the number of ionizing photons per baryon incorporated into stars $\eta_{\gamma}(M)$ as a function of stellar mass, for Pop III stars and early Pop II stars ($Z = 1/50 Z_{\odot}$).  For a given stellar track, the average ionizing photon per baryon in star is thus:
\begin{equation}
\overline{\eta}_{\gamma} = \frac{\int \eta_{\gamma}(M)\phi(M) M dM}{\int \phi(M) M dM},
\end{equation}
where $\phi(M)$ denotes the IMF.  Then, the fraction of total baryons in the universe that are in stars $F_s(z)$ must satisfy:
\begin{equation}
\frac{F_s(z_{end}) \overline{\eta}_{\gamma} f_{esc}}{C}=1,
\label{FractionOfTotalBaryonsEquation}
\end{equation}
where $z_{end}$ is the redshift at the end of Reionization, chosen to be $z_{end}=6$ in our model, and where $f_{esc}$ is the escape fraction of ionizing photons from the host galaxy into the IGM.  In the calibration of the SFR as a function of redshift, we assume the stars instantaneously produce all the ionizing photons at birth that they would normally produce during their lifetimes.  For a fixed redshift of reionization, this will underestimate the SFR.  Nevertheless, since most of the ionizing radiation was dominated by the massive stars ($M\gg 10 M_{\odot}$), with lifetimes $<10$ Myr, this is an adequate approximation.

\begin{figure}
	\begin{center}
		\includegraphics[width=0.8\columnwidth]{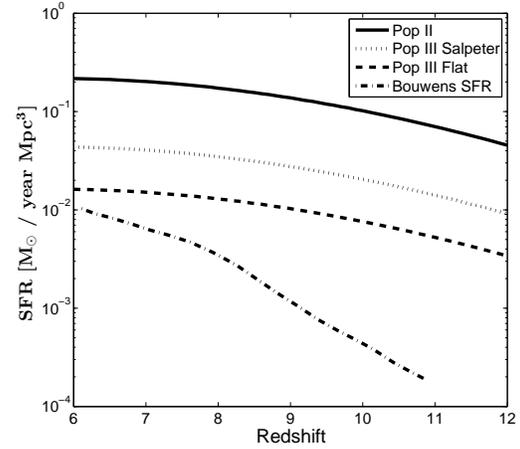}
		\caption{ The SFR for Pop II or Pop III stars required to reionize the universe by $z_{end}=6$.  The calibrated star formation rate is $f_{*}=$ 0.3\%, 0.8\%, and the unrealistically high 24.1\% for the Pop III Flat, Pop III Salpeter, and Pop II IMF models, respectively, suggesting that Pop II stars could not have driven reionization by themselves.  The SFR inferred by \citet{Bouwens2011b} from integrating the observed galaxy UV luminosity densities to $M_{AB}\approx 18$ is plotted for comparison; their substantially lower SFR is not surprising, as the contribution from the very steep faint-end slope of lower-luminosity galaxies was omitted \citep{Bouwens2011}.  The SFR for our models and the resulting SN rates all linearly scale with $C$ and $f_{esc}^{-1}$.  $C=10$ and $f_{esc} = 0.1$ were used throughout this paper.}
		\label{SFRchart}
	\end{center}
\end{figure}

The mass in stars per comoving volume as a function of redshift, $\rho_{*}(z)$, can be related to the fraction of gas in halos which converts to stars, i.e. the star formation efficiency $f_{*}$, using the Sheth-Tormen mass function $\frac{dn}{dM}$ \citep{Sheth1999} of halos:
\begin{equation}
\rho_{*}(z) = F_s(z)\rho_{b} = f_{*} \frac{\Omega_b}{\Omega_m} \int_{M_{min}}^\infty M \frac{dn(z)}{dM} dM.
\label{MassFormedIntoStarPerVolumeEq}
\end{equation}
Here $\rho_{b}$ is the cosmological baryon density, and $M_{min} \sim 10^{8} M_{\odot}$ for both Pop II and Pop III scenarios, corresponding to halos with atomic H cooling.  For cosmological parameters used in generating the Sheth-Tormen mass function, we adopt $h=0.71$, $\Omega_m=0.27$, $\Omega_{\Lambda}=0.73$, and $\Omega_b=0.045$, where $h$ is the Hubble constant in units of 100 km s$^{-1}$ and $\Omega_m$, $\Omega_{\Lambda}$, and $\Omega_b$ are the total matter, vacuum, and baryonic densities in units of the critical density \citep{Komatsu2011}.  Since $F_s(z_{end})$ is known via equation (\ref{FractionOfTotalBaryonsEquation}), by plugging $z=z_{end}$ into equation (\ref{MassFormedIntoStarPerVolumeEq}), we can calibrate the value of $f_{*}$, and evaluate $\rho_{*}(z)$ at any redshift.  Although $f_{*}$ will generally vary with redshift, here we take $f_{*}$ as a constant for simplicity of calibration.  The star formation rate is then simply:
\begin{equation}
SFR(z) = \frac{d\rho_{*}(z)}{dz}.
\end{equation}
Figure \ref{SFRchart} shows the resulting star formation rates, using $C=10$ and $f_{esc}=0.1$.  Observations of Lyman-break galaxies around $z\sim 3$ suggests that $f_{esc}$ could be larger at higher redshifts \citep{Steidel2001}.  However, theoretically the high gas densities at the redshifts of the first galaxies could decrease the escape fraction down to $f_{esc}\lesssim 0.01$ \citep{Wood2000}, in which case using $f_{esc}=0.1$ is a conservative choice that may underestimate the SFR and the corresponding SN rates.  To consider different choices of these parameters, note that the SFR in our model linearly scales with $C$ and $f_{esc}^{-1}$.

\begin{figure}
	\begin{center}
		\includegraphics[width=1.0\columnwidth]{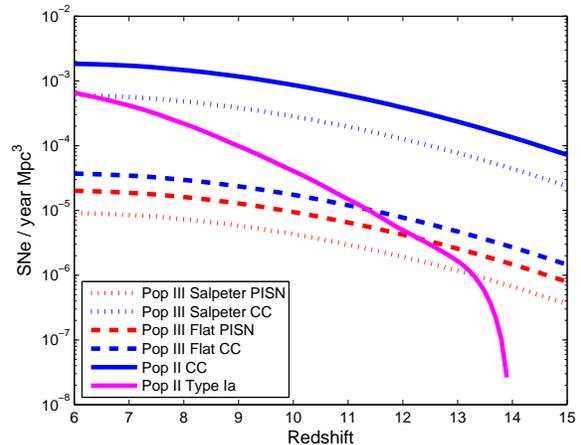}
		\caption{ The rate $R(z)$ of PISNe, CCSNe, and Type Ia SNe for our different IMF models for the stellar population responsible for reionization, per year per Mpc$^3$.}
		\label{SNRates}
	\end{center}
\end{figure}

\subsection{Snapshot Rate with JWST}

The \emph{James Webb Space Telescope}\footnote{http://www.jwst.nasa.gov/} (JWST) will include a Near Infrared Camera (NIRCam), with a spectral coverage from 0.6-5 $\mu$m with $\sim$10 nJy sensitivities in 10$^4$ s of integration time (10$\sigma$); a Near Infrared Spectrograph (NIRSpec) which operates at approximately the same wavelength range.  The Mid InfraRed Instrument (MIRI) covers 5-27 $\mu$m, but is an order of magnitude less sensitive than NIRCam.  Since isolated Pop III stars are likely beyond the reach of JWST, to test the prediction that metal-free stars had a top-heavy IMF (which has been recently debated, see \citet{Hosokawa2011}), we can either observe the cumulative properties of the first stars by imaging Pop III galaxies \citep{Zackrisson2011}, or detect their deaths as extraordinary bright supernova.

The number of new events at a given redshift that can be observed per unit solid angle is \citep{Woods1998}:
\begin{equation}
N(z) = R(z) \: (1+z)^{-1} \: \frac{dV_{c}}{dz}, \mbox{       for } z<z_{max(F;\nu)},
\label{RatePerRedshiftEquation}
\end{equation}
where $z_{max(F;\nu)}$ is the maximum redshift at which a source will appear brighter than limiting flux $F$ at an observed frequency $\nu$, $R(z)$ is the event rate per unit comoving volume, and $dV_{c}$ is the cosmology-dependent comoving volume element corresponding to a redshift interval $dz$.  The above expression includes the (1+z) reduction in apparent rate owing to cosmic time dilation.  

\begin{figure*}
\centering
\subfloat[Subfigure 1 list of figures text][Pop III Salpeter IMF model]{
\includegraphics[width=0.45\textwidth]{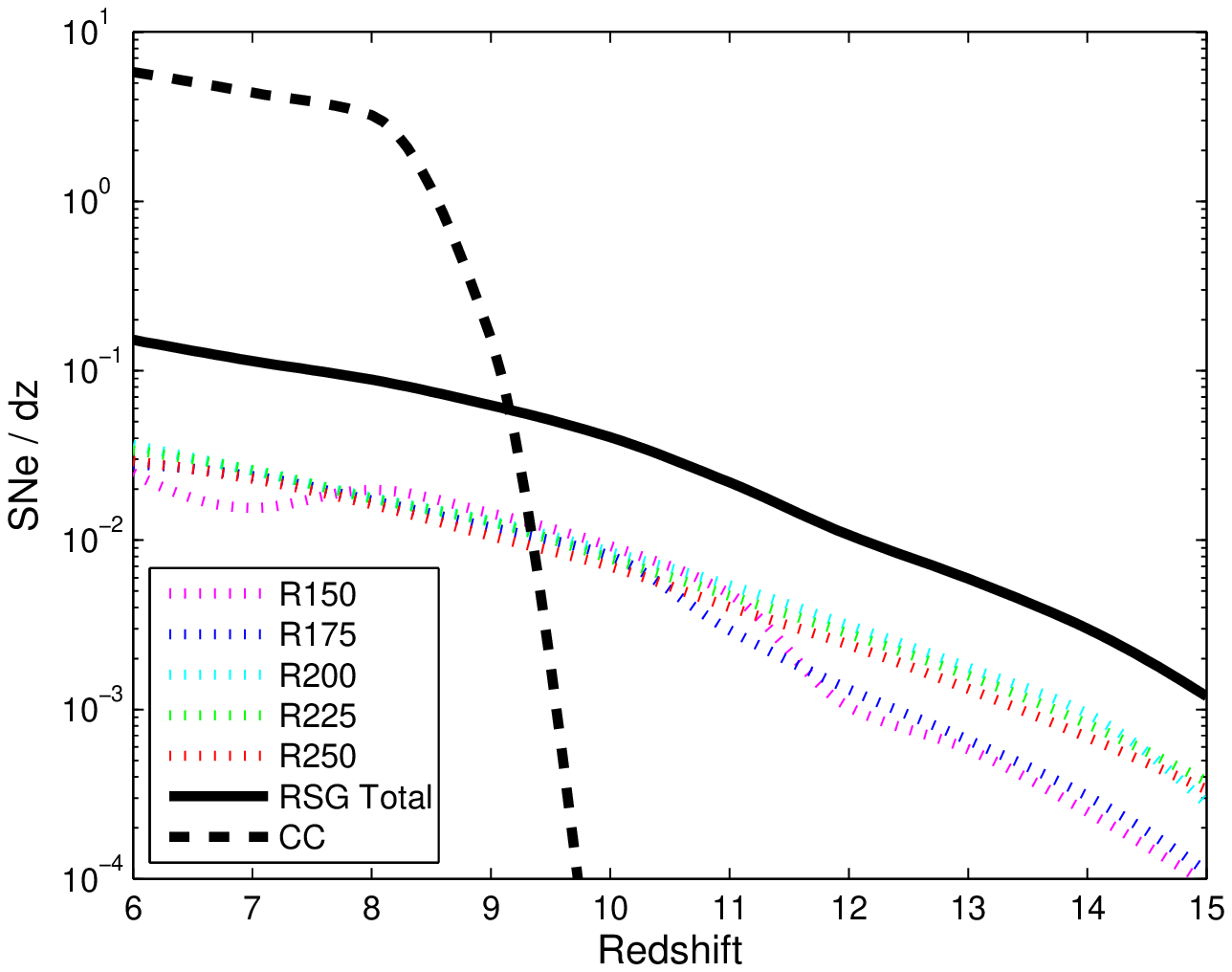}
\label{snapshot_rate_with_z:PopIIISalpeter}}
\qquad
\subfloat[Subfigure 2 list of figures text][Pop III Flat IMF model]{
\includegraphics[width=0.45\textwidth]{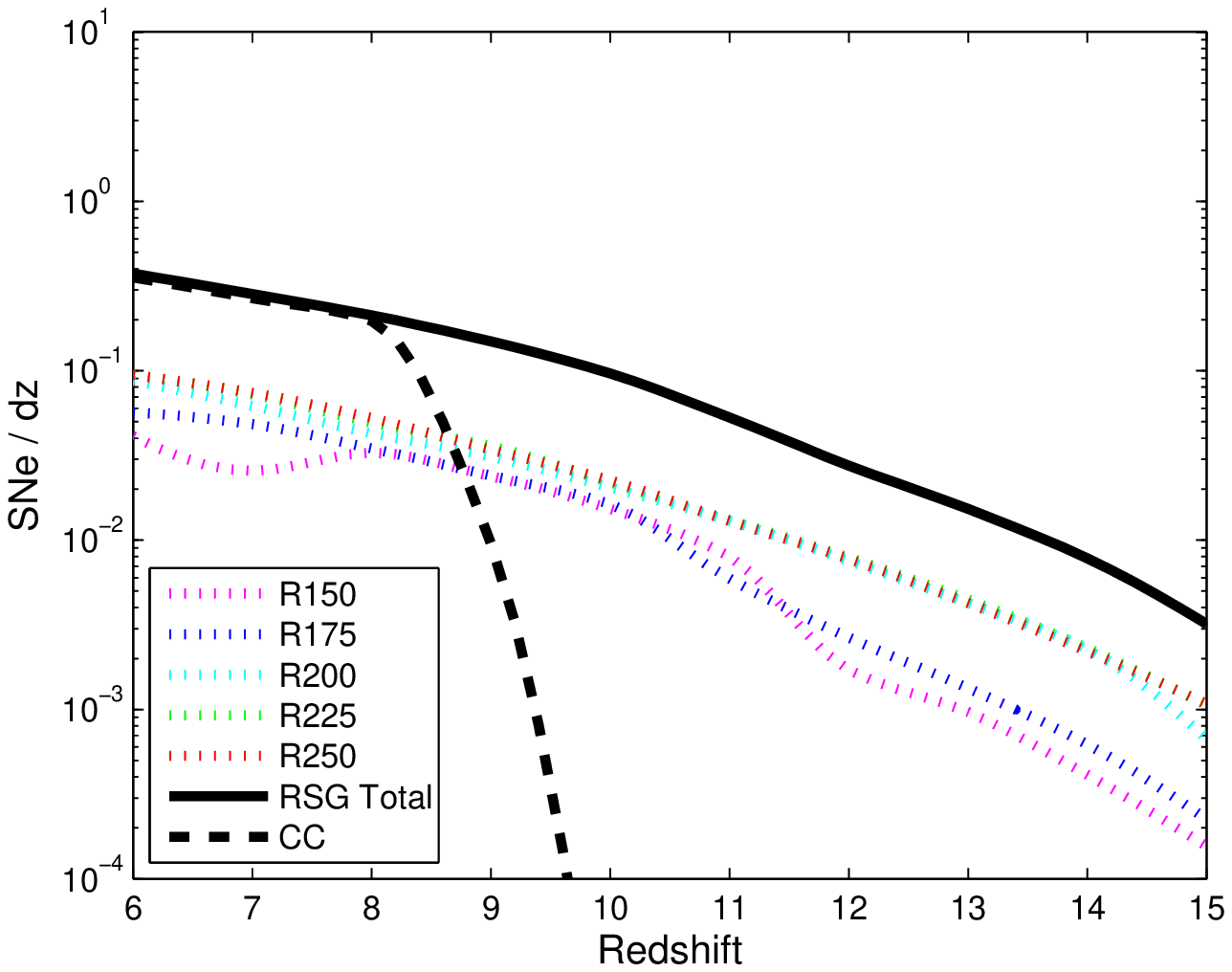}
\label{snapshot_rate_with_z:PopIIIFlat}}
\caption{ Differential snapshot rate $\frac{dN(F;\nu)}{dz}$ in the 10 arcsec$^2$ field-of-view of NIRCam on JWST, calculated using the sensitivities of the F444W filter (44400 \AA) with $t=3\times 10^4$s integration time at 5$\sigma$ (a flux threshold of 2 nJy).  At this sensitivity, each type of SN appears in rough proportion to their actual event rates (Figure \ref{SNRates}) up to $z\sim 8$, past which the ratio of detected PISNe versus CCSNe turns over, with only the brighter PISNe staying in view.  As we have not accounted for the intrinsic scatter in the luminosity of CCSNe, the actual turnover will be less sharp.  A similar turnover exists for the less massive PISN models R175, R150 starting at $z\sim 11$, past which the more massive progenitors are more likely to be seen in the field of view despite being less numerous in number given the IMF.  The snapshot rate of CCSNe in Figure \ref{snapshot_rate_with_z:PopIIISalpeter} is less than that implied in \citet{Mesinger2006}, as their assumed IMF is closer to our Pop II IMF model.}
\label{snapshot_rate_with_z:globfig}
\end{figure*}

The `snapshot rate', i.e. the total number of events (not per unit time) observed at limiting flux $F$ is:
\begin{equation}
N(F;\nu) = \int_{0}^\infty dz \: R(z) \: t(z; F; \nu) \: \frac{dV_{c}}{dz},
\label{SnapshotRateEquation}
\end{equation}
where $t(z; F; \nu)$ is the rest-frame duration over which an event will be brighter than the limiting flux $F$ at redshift $z$ for an observed frequency $\nu$.  We find this duration from our spectral time series calculated with SEDONA.  There is an implicit $(1+z)$ factor in equation (\ref{SnapshotRateEquation}) due to the time dilation of the light curve, but that cancels with the $(1+z)^{-1}$ reduction in apparent rate.  Although $t(z; F; \nu)$ of PISNe will generally be longer for more massive progenitors, the snapshot rate is not necessarily dominated by the highest mass stars, as they are less numerous, see Figure \ref{snapshot_rate_with_z:globfig}.  For CCSNe, it is not clear how the brightness of a Type IIP SN should depend on the mass of the progenitor star; here we use light curves of a $15M_{\odot}$ red giant progenitor generated with SEDONA \citep{Kasen2009}, whose broadband light curves and spectra agree very well with observed Type IIP SNe, which are observed to be the most common, at least in the nearby universe.  However, this single CCSN model means we do not capture the variation in CCSN peak flux from different progenitors, which we do so for PISNe.

\begin{table}
 	 \caption{Snapshot rate in NIRCam's field of view (10 arcsec$^2$) on JWST, using the same survey parameters as Figure \ref{snapshot_rate_with_z:globfig}.  This is the total number of PISNe and CCSNe in each NIRCam snapshot, integrated across $z \geq 6$, for the different IMF models.  One can multiply the values below by 360 to get the snapshot rate per deg$^2$.  It is not clear what fraction of PISNe explode from red supergiants (RSG) versus exposed helium cores; the rates shown in the RSG and He Core columns assume all PISN explode as that type.  Here the snapshot rate of PISNe from red supergiants is higher than the helium core model due to the longer duration of the former.  The snapshot rate for CCSNe in our Pop II IMF scenario is in good agreement with the high end estimate of 24 SNe per field found by \citet{Mesinger2006} under similar survey parameters.}
	 \begin{tabular}{ | l | l | l | l |}

    \hline
    IMF model 			& He Core 			& RSG 				& CC 					\\ \hline
    Pop III Salpeter & 0.28 				& 0.42 				& 10.43 				\\
    Pop III Flat 		& 0.74				& 1.03 				& 0.64				\\ 
	 Pop II 				& 0					& 0					& 31.83				\\ \hline
    \end{tabular}

\label{SnapshotRateNIRCam}
\end{table}

\begin{table}
	\caption{Snapshot rate in MIRI's field of view (2.35 arcsec$^2$) on JWST, using the F770W filter (77000 \AA) with $3\times 10^4$s exposure (5$\sigma$).  The resulting snapshot rate is an order of magnitude worse than NIRCam; however, these results suggest that MIRI can be used as a follow-up instrument to distinguish bright PISN events from core-collapse events.  Since MIRI is much less sensitive, the brighter He core models are more readily observable, while CCSNe cannot be seen at all in this integration time.}
	\begin{tabular}{ | l | l | l | l |}

    \hline
    IMF model 			& He Core 			& RSG 				& CC 					\\ \hline
    Pop III Salpeter & 0.05				& 0.04 				& 0 					\\ 
    Pop III Flat 		& 0.15 				& 0.12 				& 0					\\ 
	 Pop II 				& 0					& 0					& 0					\\ \hline
    \end{tabular}

\label{SnapshotRateMIRI}
\end{table}

PISNe and CCSNe occur for stars with main-sequence masses between $\sim$140-260$M_{\odot}$ \citep{Heger2002} and 8-25$M_{\odot}$ \citep{Smartt2009}, respectively.  The fate of stars between 25-140$M_{\odot}$ is uncertain; due to fallback, progenitors more massive than $\sim 40M_{\odot}$ may form black holes directly with no SN explosion \citep{Fryer1999}.  Notably, stars in the mass range 95-130$M_{\odot}$ may reach the pair production instability in the core, but the thermonuclear explosion is insufficient to unbind the star \citep{Woosley2007}, and the star undergoes pulsations of matter ejecta which may produce a very bright light curve when the shells of ejected matter collide with each other, before the star dies as a normal CCSN.  The resulting pulsation pair-instability supernova can be ultra-luminous and are presumably detectable by JWST.  However, we do not consider such events here.

Using a progenitor mass range of 8-25$M_{\odot}$ to calculate the CCSN rates is likely an underestimate; for detailed predictions on the number of CCSNe detectable by JWST at the redshifts of reionization, see \citet{Mesinger2006}, who also take into account the variation in peak magnitude of observed CCSNe and the effects of dust extinction.  For a fixed progenitor mass range, we calculate the SN rate per comoving volume $R(z)$ and find the snapshot rate shown in Tables \ref{SnapshotRateNIRCam}, \ref{SnapshotRateMIRI} and Figures \ref{SNRates}, \ref{snapshot_rate_with_z:globfig}.  In Figure \ref{snapshot_rate_with_z:globfig}, we use each red supergiant model as a proxy for the light curves of all progenitors similar in mass, e.g. R250 represents all progenitors in mass range 226-260$M_{\odot}$.

If both Pop III and Pop II stars contributed to reionization, the actual IMF will be a mixture of the Pop III and Pop II IMF used above.  By counting the number of each type of SN found in JWST snapshots, the IMF of these early stellar populations can be constrained, and the relative contribution of Pop III and Pop II stars toward reionizing the universe can be inferred.  To reduce selection effects due to the different intrinsic luminosity of the SN, a sufficiently deep exposure is needed, to enable observations of both types of SNe at the peak of their light curves should they exist at the target redshift.  

For the integration time assumed in Figure \ref{snapshot_rate_with_z:globfig}, one can directly characterize the ratio of PISNe to CCSNe  before $z\sim 8$, and set existence limits on top heavy Pop III stars up to $z\sim 10$ with a survey of $\sim 10$ JWST fields.

\subsection{Probing Intermediate Mass Stars with Type Ia SNe}

To probe the intermediate mass range ($\sim 1-8 M_{\odot}$) of the IMF during reionization, one may use Type Ia SN rates.  Type Ia supernovae (SNe Ia) are thought to occur when a white dwarf nears the Chandrasekhar mass, resulting in a thermonuclear explosion.  This requires the white dwarf to accrete mass from a binary companion.  Although the physics behind SNe Ia have been widely studied using both observations and theoretical simulations, there is still no consensus on the mechanisms that proceed the supernova.  The single degenerate model proposes the companion to be a main sequence or giant star, which donates mass via Roche lobe overflow, whereas the double degenerate scenario considers the merger of two white dwarf stars; the latter may be necessary for at least some observed Type Ia SNe \citep{Bloom2012, Schaefer2012}.  Either way, after stellar birth it takes the main sequence lifetime of the progenitor star plus an additional delay time for the Type Ia SN to proceed.

Hence, the rate of SNe Ia is empirically parametrized to follow the star formation rate (SFR), but shifted toward lower redshift after taking the delay time into account.  The SN rate at a redshift $z$ or cosmic time $t$, $R(z)=R(t)$, is given by a convolution of the SFR over delay times,
\begin{equation}
R(t) = \int_{0}^{t} SFR(t-\tau) {\rm DTD}(\tau) d\tau,
\end{equation}
where ${\rm DTD}(\tau)$ is the delay time distribution (SNe per unit time per unit stellar mass formed), in which $\tau$ is the time elapsed between the formation of the progenitor star and the explosion of the SN Ia.  Note that since the ${\rm DTD}(\tau)$ is normalized to the total stellar mass formed, it only indirectly reflects the physical efficiency of SNe Ia from their actual progenitors of $3-8 M_{\odot}$ stars \citep{Nomoto1994}. 

In previous reionization literature \citep{Haiman2009}, Type Ia SNe were expected to be extremely rare at high redshifts ($z> 6$), as the delay between the formation of the progenitor and the SN event was thought to be longer than the age of the Universe at these redshifts.  However, this view should be reconsidered in light of recent converging evidence for a prompt population of SNe Ia, see recent work by \citet{Maoz2010b}, \citet{Graur2011}, and references within.

\begin{figure}
	\begin{center}
		\includegraphics[width=1\columnwidth]{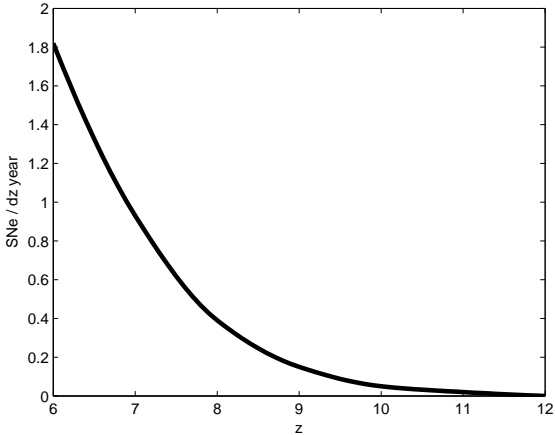}
		\caption{Type Ia SNe rate in NIRCam's field of view, per year of observation per $dz$.  For the Pop II IMF model for reionization, $\sim$1 new SNe Ia will occur every year per unit redshift at $z\sim 6-7$ in NIRCam's field of view.  The 20 days around the SN Ia peak flux at that redshift is equivalent to half a year in observer frame.  Every NIRCam snapshot of the sky, with sufficient integration time (e.g. $3\times 10^4$ s to get absolute magnitude above -18 at $z=8$ with signal-to-noise ratio of 5), will have a $\sim 50$\% probability of finding a Type Ia SN near peak flux.}
		\label{TypeIa_rate}
	\end{center}
\end{figure}

\citet{Scannapieco2005} calibrated the prompt rate via the `B' parameter, a constant of proportionality between the SFR and the prompt SN Ia rate, equivalent to the number of prompt SNe per unit stellar mass formed.  The delayed component is characterized via the parameter `A' which is the constant of proportionality between galaxy mass and the delayed SN Ia rate.  We ignore the A component as this delay exceeds the age of the universe during reionization.  The value of B is calibrated at low redshifts, for example $B=2.7-11\times 10^{-3} M_{\odot}^{-1}$ in \citet{Maoz2010}, for prompt delay times $T \in (35,330)$ Myr; here we adopt $B=3\times 10^{-3} M_{\odot}^{-1}$.  From B, we set a uniform ${\rm DTD}(\tau)=B/\Delta T$.  Using the Pop II SFR and this ${\rm DTD}(\tau)$, we calculate the event rate $R(t)$ of Type Ia SNe shown in Figures \ref{SNRates} and \ref{TypeIa_rate}.  Since the validity of these estimates depend on ${\rm DTD}(\tau)$, we assume that the astrophysics involved in shaping the forming efficiency and delay time of Type Ia SNe is not very sensitive to the cosmological epoch.

As seen in Figure \ref{TypeIa_rate}, in the scenario where Pop II stars dominated reionization, we expect $\sim$1 new SNe Ia every year per unit redshift at $z\sim 6-7$ in NIRCam's field of view of $\approx 10$ square arcseconds.  As the AB magnitude of SNe Ia during peak \citep{Hillebrandt2000} is $M_{B} \sim M_{V} \sim -19.30$ with a dispersion of 0.3, several hours integration time on JWST will be sufficient to catch a Type Ia SN near peak flux at these redshifts.

Regardless of which population of stars dominated reionization, Type Ia SNe offers a way to probe the intermediate mass range of the reionization IMF.  In addition, the $z\gtrsim 6$ sky offers an unambiguous way of isolating a prompt population of SNe Ia, as the universe was not sufficiently old in the epoch of reionization for the delayed component of Type Ia SNe to contribute any events.  Therefore, Type Ia SNe at $z>6$ could be used to test whether the delay times are indeed connected to SN Ia formation mechanisms and properties.  Finally, the existence of standard candles during the epoch of reionization could be useful for cosmological measurements.  In the Appendix, we discuss that probing reionization history with Type Ia SNe by constraining the global ionization fraction using Thomson optical depth measurements requires an unrealistic survey time for JWST.

\subsection{Typing the Supernovae}

At lower redshifts, SNe are usually typed by spectral lines.  Furthermore, the smoking gun evidence for a PISN is the measurement of a large core ($>50M_{\odot}$) composed of helium or other heavier elements.  In the case of SN 2007bi, \citet{Gal-Yam2009} analyzed the nebular spectrum 16 months after peak light to infer $\sim 4M_{\odot}$ of $^{56}$Ni, implying a large core mass $>50M_{\odot}$ as in a pair-instability explosion (but see \citet{Moriya2010} for a CCSN model for SN 2007bi that ejects $6.1 M_{\odot}$ of $^{56}$Ni).  However, spectroscopic typing of high redshift SNe seen by JWST may be unrealistic; for example, at around  3 $\mu$m on NIRSpec, achieving a signal-to-noise ratio of 5 at redshift $\sim 8$ would require one full day of integration time.  Alternatively, in anticipation of Pan-STARRS\footnote{http://pan-starrs.ifa.hawaii.edu/public} and LSST\footnote{http://www.lsst.org/lsst/} increasing the number of photometrically detected SNe to a few hundred thousand in the next two decades, much work has been done in the photometric identification and classification of SNe \citep{Kessler2010}.  As photometric classification of SNe matures, it could complement or replace the spectroscopic typing of high-z SNe, reducing the required JWST time.

The light curves of the more massive PISNe are very luminous ($10^{43}$-$10^{44}$ erg s$^{-1}$) and long-lasting ($\sim 300$ days), characteristics that do not exist for most other types of SNe.  As long as the SN redshift is known, multi-epoch observations can determine its rest-frame luminosity and duration, and identify the more massive PISNe explosions.  Aside from the most energetic events, typing PISNe using their magnitude and color will be difficult.  Despite their enormous kinetic energies of $\sim 10^{53}$ ergs, the peak optical luminosities of PISNe are similar to those of other SNe, even falling below the Ia and II curves for smaller mass progenitors.  The majority of PISNe will actually be these dimmer events.  Also, since PISNe spend most of their lives in the same temperature range as other SNe, their colors are also similar.

An extended light curve, rather than an extreme luminosity or unusual color, may therefore be the most important signature of PISNe.  In particular, the distinguishing feature of PISNe is its exceptionally long rise time, $\gg 100$ days in the rest frame.  Also, the detection of a slow decline rate that follows the decay rate ($\sim$0.01 mag/day) of $^{56}$Co, the product of $^{56}$Ni decay, would provide strong evidence that the SN synthesized significant amounts of $^{56}$Ni.  At $z=8$, even with time dilation, this results in $\sim$0.4 mag variation per year, which should be within the sensitivity of a multi-year JWST survey; the decline in the bluer bands are 2-3 times larger, due to the onset of iron group line blanketing.  The detection of a secondary maximum in the light curve also supports the synthesis of $^{56}$Ni.  However, the lack of a secondary peak does not rule out a large presence of $^{56}$Ni, as strong radial mixing could smear out the two bumps \citep{Kasen2006a}.

\subsection{Survey Strategies}

The long duration of high redshift PISN light curves, prolonged by cosmological time dilation, poses a great challenge for detecting them as transients.  At z$\sim$7 the light curve of a PISN can last for over 1000 days in the observer's frame.  Without spectroscopic measurements, the telltale sign of a massive progenitor PISN is an incredibly long plateau in its light curve.  Therefore, instead of a threshold experiment, we suggest a search strategy that involves taking a series of `snapshots' of a field, each snapshot separated by $\sim 1$ year, and searching for variations in the flux of objects in successive images.  Since Pop III star formation occurs in the smallest galaxies, blank-field surveys should be the sufficient for searching for PISNe.

The total number of SNe detected in a survey of total integration time $t_{surv}$ is
\begin{equation}
N_{surv} = \frac{1}{2}\frac{t_{surv}}{t_{exp}} \frac{\Delta \Omega_{FOV}}{4\pi} N_{exp},
\end{equation}
where $\Delta \Omega_{FOV}$ is the instrument's field of view, $t_{surv}/t_{exp}$ is the number of fields which
can be tiled within the survey time $t_{surv}$, and $N_{exp}$ is the snapshot rate from equation (\ref{SnapshotRateEquation}), i.e. the number of SNe bright enough to be detected in an exposure of duration $t_{exp}$ \citep{Haiman2009}.  The factor of $\frac{1}{2}$ is included to account for observations in $4$ color bands (2 pairs of filters, as NIRCam observes in two bands simultaneously using a dichroic) for determining photometric redshift and typing of the SNe.  To detect SNe by their variability, each field requires repeated observations, and therefore any survey should piggyback on fields that have already been observed.

In the case where several fields are already available from other JWST surveys, the snapshot rates given by Table \ref{SnapshotRateNIRCam} suggest that a dedicated, long program may not be required to detect dozens of high redshift SNe.  Due to the order-of-magnitude difference in the snapshot rate of PISNe vs CCSNe for the different IMF models, more than $10$ fields with followup repeated imaging should already help constrain the stellar population responsible for reionization.  Cosmic variance will affect the total number of SNe for small number of fields, but the ratio of PISNe to CCSNe would still be indicative of the IMF.  For example, if Pop III (Flat IMF model) and Pop II stars had equal contribution to reionization (which means Pop II stars dominate Pop III stars by roughly 20-to-1 in total mass), one could use 20 images conducted for other programs as references and only revisit the same image twice for a total of 3 snapshots per field of view.  Observing in 4 bands, for a total of 28 days integration time over 2 years, such a survey expects to see $\sim$10 red supergiant PISNe and $\sim$300 CCSNe.

To the extent that PISN spectra can be represented as a distribution of blackbodies at different temperatures, since the temperature and redshift would be degenerate, it will be impossible to acquire photometric redshifts without further information about the SN epoch.  However, our simulated spectra show significant deviations from a blackbody in the UV ($l < 3500$\AA) due to metal-line blanketing in the SN photosphere, providing spectral and photometric signatures that could be used as redshift indicators, depending on their strength.

\subsection{Luminosity Function}

\begin{figure}
\centering
		\includegraphics[width=1\columnwidth]{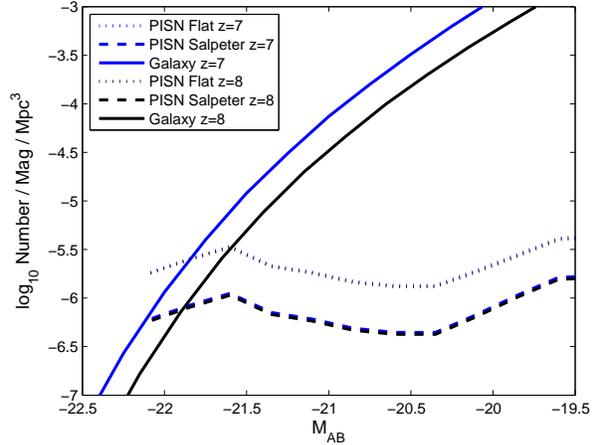} \\
	\caption{Luminosity functions of PISNe at 4000\AA for $z=7$ and $8$, for the Pop III Flat IMF or Salpeter IMF star formation rate models, compared with the galaxy luminosity function.  For the Pop III Flat model, the volumetric count of PISNe exceeds galaxies past $M_{AB} \sim -21.5$; the dominance of PISNe should become greater at higher redshifts, as galaxies decrease in luminosity while PISNe stay the same.  The PISN luminosity functions at $z=7,$ $8$ overlaps coincidentally because the increase in time dilation compensates for the decrease in PISN event rate.}
\label{luminosity_function}
\end{figure}

Although the UV flux of PISNe is relatively short lived, the more massive PISNe stay bright in its rest frame visible band for over a year.  Given this brightness and long intrinsic duration, coupled with the $(1+z)$ time dilation at high redshifts, it is conceivable that PISNe could contribute to the luminosity function of all objects at high redshifts when galaxies were dim.  Figure \ref{luminosity_function} illustrates the luminosity function of PISNe at $\sim$4000\AA, calculated using the helium core progenitor models for PISN luminosity, and the Pop III Flat or Pop III Salpeter models for the star formation rate.  Shown for comparison are the projected galaxy luminosity functions at high redshifts, using the \citet{Bouwens2011b} best fit Schechter parameterization for the UV luminosity function, and shifting to the visible band using $U-V \approx 0.4$, $0.3$ for $z=7,$ $8$ respectively, measured using the Spitzer Infrared Array Camera \citep{Labb'e2010, Labb'e2010a}.  Applying this U-V shift is a crude approximation, as luminous and faint galaxies have different rest frame UV-to-optical color; however, we are most interested in the bright end of the luminosity function, where this current U-V measurement is applicable.

The luminosity function for PISNe implied by our Pop III IMF models overlaps with the galaxy luminosity function at the brightest magnitudes.  If a top-heavy Pop III IMF was solely responsible for reionization, PISNe will contaminate the brightest end of the galaxy luminosity function, unless great care is taken to remove these supernovae.  Since the volumetric count of the brightest galaxies and PISNe is very low, it will take a wide infrared survey to observe this effect.

\section{Discussion}

In our discussion we ignored complicating factors such as metallicity and rotation, and calculated the PISN and CCSN event rate using only the SN progenitor mass range along with the star formation rate.  However, at low redshifts $z<1$, the measured CCSN rate is a factor of $\sim 2$ smaller than that predicted by the analogous calculation using the measured cosmic star formation rate.  The discrepancy is likely due to many intrinsically low-luminosity or obscured SNe being missed in surveys \citep{Horiuchi2011}.  As this discrepancy is lower than the uncertainty in our SFR model parameters, and we already account for lower intrinsic luminosities for the lower progenitor mass PISNe, we do not take obscuration into account for our predictions of the SN rate as seen by JWST. 

The IMF of early stellar populations responsible for reionization should also leave an imprint on the metal enrichment pattern via their SN products.  So far, the abundance patterns observed to date in extremely metal-deficient stars in the Galactic halo \citep{Beers2005} are more consistent with an IMF that produced much more CCSNe instead of PISNe \citep{Joggerst2010}.  However, in previous surveys, subtle selection effects might have disfavored finding PISN-enriched stars; the metal yields of PISNe are so high that the metal abundances of stars formed out of PISN ejecta \citep{Greif2008} are already higher than the metallicity range targeted by metal-deficient star surveys \citep{Karlsson2008}.  

Large carbon enhancements observed in metal-poor stars, when interpreted as the outcome of pollution by winds from binary companions that have gone through the AGB phase, suggest the existence of a large number of intermediate-mass stars ($\sim 1-8 M_{\odot}$) at high redshifts \citep{Tumlinson2007,Tumlinson2007a}.  Alternatively, nucleosynthesis in faint CCSNe from higher mass stars could also explain the observed carbon enhancement in metal-poor stars \citep{Iwamoto2005}.  Observing the Type Ia SN rate during the epoch of reionization will be an complementary way to test these models, and constrain the number of intermediate-mass stars at high redshifts.

The predicted initial mass range of $\sim$140 to 260 $M_{\odot}$ for PISN progenitors assumed the stars to be non-rotating \citep{Heger2002}.  However, observations find that at very low metallicities, stars rotate faster \citep{Martayan2007}.   The fast rotation of the first stars is supported by the latest hydrodynamic simulations of their formation \citep{Stacy2011}, and also by observations of anomalously high abundances of Ba and La with respect to Fe in ancient low-mass stars \citep{Chiappini2011}, which could originate in metal-poor fast-rotating massive stars.  Generally, rotation should increase the required PISN progenitor mass by increasing mass loss.  \citet{Meynet2006} found that, contrary to the usual $\dot{M}\propto Z^{0.5}$ scaling relation, rotating stars at very low metallicity $Z \sim 10^{-5}$ to $10^{-8} Z_{\odot}$ show a large mass loss, up to $\sim 50\%$, mainly resulting from efficient mixing of stellar nucleosynthesis products into the stellar surface.  However, \citet{Ekstrom2008} found that for strictly $Z=0$ stars, the mass loss is very low, even for models that reach critical velocity in the main sequence.  These results imply that, for rapidly rotating Pop III stars to die as a PISN, the required progenitor mass is extremely sensitive to whether the star is truly metal-free or not.

At much lower redshifts, PISNe have likely already been observed, most persuasively in the case of the very luminous and long duration event SN 2007bi \citep{Gal-Yam2009}.  Other more recent candidates include PTF 10nmn (Gal-Yam in preparation; Yaron et al. in preparation) and PS1-11ap  (Rubina Kotak et al. in preparation).  As pristine gas was recently observed at redshifts after reionization \citep{Fumagalli2011}, it is possible that some low redshift PISNe have Pop III progenitors born out of surviving pockets of metal-free gas; the rates of PISNe in this scenario was considered by \citet{Scannapieco2005a}.  However, the metallicities of the host galaxies of SN 2007bi and PTF 10nmn are well above the metallicity threshold required to form Pop III stars \citep{Young2010}.  Therefore, it is plausible that PISNe can have very massive Pop II/I progenitors as well, perhaps born via the merger of stars in collision runaways in young, dense star clusters \citep{Pan2011}.

\section{Conclusions}

We analyzed simulated light curves and spectra of pair-instability supernovae for a variety of progenitor masses and envelope types, and found that the supernovae from the more massive progenitors are super-luminous and have extended light curves,  traits that would help  photometrically distinguish pair-instability supernovae from other types of supernovae using repeated snapshots.  We calculated the rates and detectability of pair-instability, core collapse, and Type Ia supernovae during the redshifts of reionization, and showed that it is possible to constrain the initial mass function of stars at that time, and identify the stellar population responsible for reionization.  If Pop III stars made the dominant contribution of ionizing photons during reionization, the bright end of the galaxy luminosity function will be contaminated by pair-instability supernovae.

\bigskip
\section*{Acknowledgments.}
We thank Bob Kirshner, Kaisey Mandel, and Jonathan Pritchard for helpful discussions.  TP was supported by the Hertz Foundation.  This work was supported in part by NSF grant AST-0907890 and NASA grants NNX08AL43G and NNA09DB30A.  This work is supported  by the Director, Office of Energy
Research, Office of High Energy and Nuclear Physics, Divisions of Nuclear Physics, of the U.S. Department of Energy under Contract No.DE-AC02-05CH11231. This research has been supported by the DOE SciDAC Program (DE-FC02-06ER41438).  We are grateful for computer time provided by ORNL through an INCITE award and by NERSC.

\bibliographystyle{mn2e}
\bibliography{references}

\clearpage

\appendix

\section{On the Difficulty of Mapping Reionization History with Type Ia SNe}

\subsection{Constraining the Ionization Fraction}

Barring some extreme evolution of the IMF, the neutral fraction of the IGM is expected to rise quickly toward high redshift, with the mean neutral fraction of the IGM expected to reach 6-12$\%$ at z=6.5, 13-27$\%$ at z=7.7 and 22-38$\%$ at z=8.8 \citep{Cen2010}.  Currently, the most stringent observational probe on the ionization history of the IGM is the total Thomson optical depth seen by WMAP, $\tau = 0.088 \pm 0.015$ \citep{Komatsu2011}.  The Planck mission is projected to reduce the error bars to $0.01$.  However, $\tau$ cannot break degeneracies between different reionization histories; for example, both a rapid, early reionization or an extended, late reionization may have the same total Thomson optical depth.  Finding $\tau(z)$ using Type Ia supernovae at high redshifts would break this degeneracy.

We set up a toy model of the global average ionization fraction $X(z)$ using the Fermi-Dirac form for the ionization fraction:
\begin{equation}
X(z) = \frac{1}{e^{\frac{z-z_{re}}{\Delta}}+1},
\end{equation}
where $z_{re}$, $\Delta$ are model parameters that characterize the redshift and duration of reionization, respectively.  Then for an luminous object at redshift $z_{obs}$, the Thomson electron scattering optical depth is the integral of $X(z) n_{e} \sigma_{T}$, the ionization fraction times the electron density times the Thomson cross section integrated along proper length \citep{Shull2008},
\begin{eqnarray}
\tau(z_{obs}) &=& \int_{0}^{z_{obs}} X(z) n_{e} \sigma_{T} \frac{c}{(1+z)H(z)} dz \\
              &+& \tau_{HeIII}\times \Theta(z_{obs}-z_{HeIII}).
\end{eqnarray}
Here $H(z)=H_0[\Omega_{m}(1+z)^3+\Omega_{\Lambda}]^{1/2}$.  The second term with $\tau_{HeIII} \simeq 0.002$, and $\Theta$ as the Heaviside step function, comes from the full reionization of HeII to HeIII around $z_{HeIII} \sim 3$ \citep{Shull2004}.  

\begin{figure}
	\begin{center}
		\includegraphics[width=0.9\columnwidth]{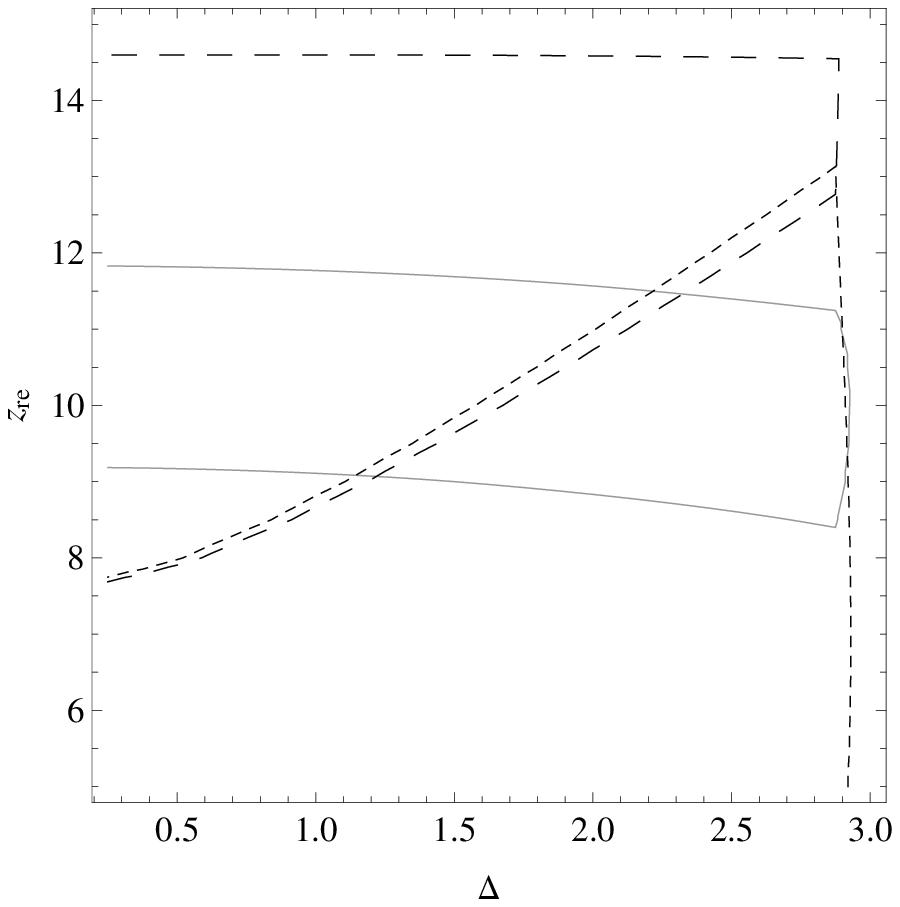}
		\caption{One-sigma (68\%) contours of $z_{re}$, $\Delta$ given measurements of $\tau(z_{obs}=8)$ and CMB total optical depth.  Flat priors of $z_{re}=5-15$, $\Delta=0.25-3$ are assumed for the x,y axis, respectively.  The straight, long-dashed, and short-dashed line represent the contours from measurements of $\tau_{CMB}=0.088 \pm 0.015$ \citep{Komatsu2011}, $\tau(z_{obs}=8)=0.07 \pm 0.01$, and $\tau(z_{obs}=8)=0.04 \pm 0.01$, respectively, the latter two of which are hypothetical values of optical depth we may measure from a large sample of SNe Ia.}
	\label{ContourOneSigma_zObs8_deltatau0.01}
	\end{center}
\end{figure}

Using this model with hypothesized optical depth measurements, we use Bayesian methods to find the corresponding probability distribution for our reionization history model parameters $z_{re}$, $\Delta$ shown in Figure \ref{ContourOneSigma_zObs8_deltatau0.01}.  These measurements along with the known optical depth to the CMB can help constrain the duration of reionization $\Delta$.

\subsection{Survey Feasibility with JWST}

Using the Phillips relation \citep{Phillips1993}, we can utilize Type Ia SNe as standard candles by measuring the shape of the light curve.  Specifically, to characterize the light curve and find $\Delta m_{15}(B)$, i.e., the decline in the B-magnitude light curve from maximum light to the magnitude 15 days after B-maximum, we should to take a snapshot every 3 days for roughly 20 days (5 days before and 15 days after peak), which at redshift 8 means returning to the same field of view once every month due to time dilation.  Note that above $z>10$, the B band ($\sim 4400 \text{\AA}$) redshifts out of NIRCam.  Also, \citet{Brandt2010} found that the prompt channel Type Ia SNe are more luminous (high-stretch, slow declining), and thus have a lower $\Delta m_{15}(B)$.  For NIRCam, we can see up to -18 AB magnitude at redshift $\sim 8$ with $3\times 10^4$ s integration time ($\sim 8$ hours), with a signal-to-noise ratio of 5, and fully characterize the SN light curve as long as $\Delta m_{15}(B) < 1.3$.  This is roughly $\sim$60 hours of total integration time over half a year for each field of view with a Type Ia SN.  Unrealistically assuming command of all JWST's time, the light curve of $\sim$150 Type Ia per year can be fully mapped out.

As seen in Figure \ref{TypeIa_rate}, JWST might see $\sim$1 Type Ia SNe at $z=6-7$ for every snapshot it takes with NIRCam.  To find interesting results about reionization history, one should probe the ionization fraction before the end of reionization, at $z\sim 8$ or above.  Here, the SNe are dimmer and the rate is smaller, though by only a factor of $\sim 2$ each.  A bigger obstacle is, at this redshift, many color bands redshift out of NIRCam's range (at $z\sim 8$, only the UBV bands at the SN rest frame are still accessible).  Since the calibration of Type Ia SNe magnitudes relies on multiple color bands, it is not clear the often quoted $\sim 0.20$ standard deviation in distance modulus can be achieved.  Moreover, calibrating the intrinsic luminosity of SNe Ia down to $\Delta M \sim 0.20$ is equivalent to a variation in optical depth of $\Delta \tau \sim 0.18$.  To get the Thomson optical depth to $1\%$ precision at a fixed redshift bin, required to make meaningful constraints on reionization history, would require over 300 independent Type Ia SNe, or over two full years of JWST's integration time.  This is clearly not feasible.

\end{document}